\documentclass[modern,trackchanges]{aastex631}
\usepackage{amsmath}

\begin{document}

\title{\texttt{exoatlas}: friendly Python code for exoplanet populations}
\shorttitle{Probabilistic Cosmic Shorelines}

\author[0000-0002-3321-4924]{Zach K. Berta-Thompson}
\affiliation{University of Colorado Boulder, Department of Astrophysical and Planetary Sciences}
\email{zach.bertathompson@colorado.edu}  

\author[0000-0001-6484-7559]{Patcharapol Wachiraphan}
\affiliation{University of Colorado Boulder, Department of Astrophysical and Planetary Sciences}

\author[0009-0009-7798-8700]{ Autumn Stephens}
\affiliation{University of Colorado Boulder, Department of Astrophysical and Planetary Sciences}

\author[0009-0000-9058-0069]{Mirielle Caradonna}
\affiliation{University of Colorado Boulder, Department of Astrophysical and Planetary Sciences}

\author[0000-0001-8504-5862]{Catriona Murray}
\affiliation{University of Colorado Boulder, Department of Astrophysical and Planetary Sciences}

\author[0009-0007-0158-1651]{Valerie Arriero}
\affiliation{University of Colorado Boulder, Department of Astrophysical and Planetary Sciences}

\author[0009-0003-0287-0149]{Jackson Avery}
\affiliation{University of Colorado Boulder, Department of Astrophysical and Planetary Sciences}

\author[0009-0009-7798-8700]{Autumn Stephens}
\affiliation{University of Colorado Boulder, Department of Astrophysical and Planetary Sciences}

\author[0000-0002-7119-2543]{ Girish M. Duvvuri}
\affiliation{Department of Physics and Astronomy, Vanderbilt University}

\author[0000-0002-4489-0135]{Sebastian Pineda}
\affiliation{Laboratory for Atmospheric and Space Physics, University of Colorado Boulder}

\section{Summary}
\label{summary}

Planets are complicated. Understanding how they work requires connecting
individual objects to the context of broader populations. Exoplanets are
easier to picture next to their closest Solar System archetypes, and
planets in the Solar System are richer when seen alongside a growing
community of known exoplanets in the Milky Way. The \texttt{exoatlas}
toolkit provides a friendly Python interface for retrieving and working
with populations of planets, aiming to simplify the process of placing
worlds in context.

\section{Statement of need}
\label{statement-of-need}

\texttt{exoatlas} was designed to meet a need among both researchers and
educators for an intuitive Python tool to access planet populations.
Particularly in working with students and junior scientists, for whom
easy avenues for exploration and play would have particular benefit,
significant barriers often frustrate attempts to perform the following
tasks:

\begin{itemize}
\item
  retrieving basic properties for exoplanets + Solar System objects
\item
  calculating derived planet quantities with propagated uncertainties
\item
  comparing individual exoplanets to relevant comparison samples
\item
  planning future telescope observations of known exoplanet systems
\item
  making beautiful and up-to-date planet data visualizations
\end{itemize}

Online planetary data archives merge incredible curatorial efforts with
powerful tools for data access and visualization, including the NASA
Exoplanet Archive \citep{exo-archive}, exoplanet.eu \citep{exoplaneteu},
exo.MAST \citep{exomast}, TEPCat \citep{tepcat}, Open Exoplanet Catalog
\citep{open-exo} for exoplanets and the JPL Solar System Dynamics
\citep{ssd}, IAU Minor Planet Center \citep{mpc}, the NSSDCA Planetary
Facts sheets \citep{facts} for Solar System objects. \texttt{exoatlas}
does not intend to replace any of these important archival efforts (it
pulls nearly all its data from them); rather, \texttt{exoatlas} aims to
provide an approachable interface for exploratory analysis and
illuminating visualizations to help the community make better use of
these resources.

\section{\texorpdfstring{Mapping populations with
\texttt{exoatlas}}{Mapping populations with exoatlas}}\label{mapping-populations-with-exoatlas}

The user interface for \texttt{exoatlas} centers around the Python
\texttt{Population} class, with each \texttt{Population} object
containing a standardized table of planet properties and methods for
interacting with that table. \texttt{exoatlas} makes extensive use of
\texttt{astropy} \citep{astropy} to be as familiar as possible for modern
astronomers, and it is thoroughly documented with astronomical audiences
in mind at
\href{https://zkbt.github.io/exoatlas/}{zkbt.github.io/exoatlas/}.

\begin{figure}
\centering
\includegraphics[width=\textwidth]{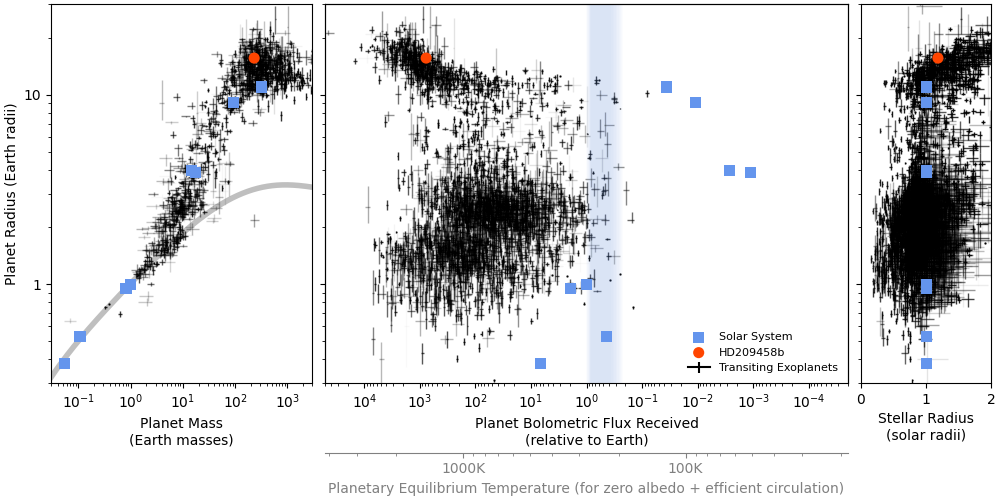}
\caption{Example \texttt{exoatlas} visualization placing the first
discovered transiting exoplanet HD209458b in context with other
transiting exoplanets and the eight major Solar System planets.
Errorbars use a color intensity that scales inversely with quantity
uncertainties, to avoid giving undue visual weight to the least precise
data. \label{fig:exoatlas}}
\end{figure}

\begin{itemize}

\item
  To retrieve planet data, \texttt{exoatlas} provides
  \texttt{Population} objects that automatically access archive data for
  exoplanets as well as Solar System major planets, minor planets, and
  moons. Exoplanet data come from the NASA Exoplanet Archive
  \citep{exo-archive} API, and Solar System data come from the JPL Solar
  System Dynamics \citep{ssd} API or small reformatted data tables
  included in the package repository itself. Whatever the original data
  source, all \texttt{Population} objects act similarly and have uniform
  nomenclature for accessing data columns. Planet quantities all have
  physical units attached with \texttt{astropy.units} to facilitate unit
  conversions and minimize conceptual errors. New \texttt{Population}
  objects can also be created from \texttt{astropy.table} tables,
  enabling custom datasets to be included.
\item
  To calculate derived quantities for planets, a set of default methods
  are included within the core \texttt{Population} definition, or users
  may attach their own calculation methods. If data for a quantity is
  missing, calculations can be used to swap in alternate estimates; for
  example, a planet's semimajor axis \(a\) will attempt first to pull
  from the original data table, and then second to calculate \(a\) from
  the planet's period \(P\) and the star's mass \(M_\star\) assuming
  Newton's Version of Kepler's Third Law \(P^2 = 4\pi^2 a^3/GM_\star\),
  and then third to calculate from a transit-derived ratio \(a/R_*\).
  Uncertainties on all derived quantities can be numerically propagated
  using the \texttt{astropy.uncertainty} framework, where distributions
  of samples are generated for each original table quantity, carried
  through calculations, and then used to estimate confidence intervals.
\item
  To extract sets of planets meeting particular criteria,
  \texttt{Population} objects can be indexed, sliced, and masked to
  generate new smaller \texttt{Population} objects. Coupling familiar
  array or table operations into the creation of subpopulations enables
  nuanced filtering of datasets based on any combination of original
  archival table quantities, derived quantities, and/or quantity
  uncertainties.
\item
  To plan telescope observations of known exoplanet systems,
  \texttt{exoatlas} can estimate the signal-to-noise (S/N) ratio
  acheivable for exoplanet observables. As derived quantities, all S/N
  estimates can include propagated uncertainties, to help avoid biasing
  target samples toward systems with larger uncertainties. Using
  \texttt{astroplan} \citep{astroplan} and \texttt{astropy.coordinates}
  \citep{astropy}, \texttt{exoatlas} can determine local altitudes and
  azimuths for all elements in a \texttt{Population}, as well as
  upcoming opportunities to observe transits from ground-based
  telescopes (when the star is above the horizon, the Sun is down, and
  the planet is passing in front of the star). Including basic
  observation planning tools enables a workflow where targets can be
  filtered both by the estimated detectability of a signal and by
  whether a telescope can actually point at the system.
\item
  To make explanatory illustrations of planet data, \texttt{exoatlas}
  includes a visual language to build up population comparison plots.
  The core elements of this visual language are the \texttt{Plottable}
  (a quantity that should be represented with certain scaling, limits,
  and labels), the \texttt{Map} (a panel expressing plottable quantities
  with position, size, or color), and the \texttt{Gallery} (a collection
  of maps with linked axes and/or datasets). Preset visualizations built
  from this language can offer quick contextual reference for a planet's
  fundamental properties, as in Figure \autoref{fig:exoatlas}, which was
  generated with only six lines of code:
\end{itemize}

\begin{verbatim}
            from exoatlas import *
            from exoatlas.visualizations import *

            e = TransitingExoplanets()
            s = SolarSystem()
            h = e["HD209458b"]

            PlanetGallery().build([e, s, h])
\end{verbatim}

\section{Research, teaching, and
learning}\label{research-teaching-and-learning}

\texttt{exoatlas} was designed to support the researcher who wants to
contextualize planet populations in papers, proposals, or talks, the
educator who wants help connect lessons more immediately to real
exoplanet and Solar System data, and the student who simply wants to
learn a little more about oodles of neat planets. All these communities
are encouraged try \texttt{exoatlas}, to ask for help, to suggest
improvements, and/or to contribute code!

\section{Acknowledgements}\label{acknowledgements}

We acknowledge the long commitment federally-funded archives have made
to preserving and sharing data with the scientific community, and the
heroic efforts of the people who build, maintain, and continually
improve those archives. This material is based upon work supported by
the National Science Foundation under Grant No.~1945633.

\bibliography{paper}{}
\bibliographystyle{aasjournalv7}

\end{document}